\documentclass[twocolumn]{aastex63}
\usepackage{savesym}
\savesymbol{tablenum}
\usepackage{siunitx}
\restoresymbol{SIX}{tablenum}

\usepackage[utf8]{inputenc}
\usepackage{graphicx}
\usepackage{amsmath}
\usepackage{siunitx,booktabs}
\usepackage{array}
\usepackage{longtable}
\usepackage{lineno}

\newcommand{\oii}{[O \textsc{ii}]$\lambda3727$}

\newcommand{\oiii}{[O \textsc{iii}]$\lambda5007$}
\newcommand{\nii}{[N \textsc{ii}]$\lambda6583$}
\newcommand{\hb}{H$\beta$}
\newcommand{\hg}{H$\gamma$}
\newcommand{\ha}{H$\alpha$}

\newcommand{\numgal}{145 }
\usepackage{xspace}

\shorttitle{LEGA-C Mass--Metallicity Relation}

\shortauthors{Lewis et al.}

\begin{document}

\title{The Gas-Phase Mass--Metallicity Relation for Massive Galaxies at $z\sim0.7$ with the LEGA-C Survey}

\author{Zach J. Lewis}
\affiliation{Department of Physics and Astronomy and PITT PACC, University of Pittsburgh, Pittsburgh, PA 15260, USA}
\affiliation{Department of Astronomy, University of Wisconsin-Madison, Madison, WI, 53706}

\author[0000-0001-8085-5890]{Brett H.\ Andrews}
\affiliation{Department of Physics and Astronomy and PITT PACC, University of Pittsburgh, Pittsburgh, PA 15260, USA}

\author[0000-0001-5063-8254]{Rachel Bezanson}
\affiliation{Department of Physics and Astronomy and PITT PACC, University of Pittsburgh, Pittsburgh, PA 15260, USA}

\author{Michael Maseda}
\affiliation{Department of Astronomy, University of Wisconsin-Madison, Madison, WI, 53706}

\author[0000-0002-5564-9873]{Eric F. Bell}
\affil{Department of Astronomy, University of Michigan, 1085 South University Avenue, Ann Arbor, MI 48109-1107, USA}

\author{Romeel Dav\'e}
\affil{Institute for Astronomy, Royal Observatory, University of Edinburgh, Edinburgh EH9 3HJ, UK}
\affil{Department of Physics \& Astronomy, University of the Western Cape, Robert Sobukwe Rd, Bellville, 7535, South Africa}

\author[0000-0003-2388-8172]{Francesco D'Eugenio}
\affil{Kavli Institute for Cosmology, University of Cambridge, Madingley Road, Cambridge, CB3 0HA, UK}
\affil{Cavendish Laboratory, University of Cambridge, 19 JJ Thomson Avenue, Cambridge, CB3 0HE, UK}

\author[0000-0002-8871-3026]{Marijn Franx}
\affil{Leiden Observatory, Leiden University, P.O. Box 9513, 2300 RA, Leiden, The Netherlands}

\author[0000-0002-9656-1800]{Anna Gallazzi}
\affil{INAF-Osservatorio Astrofisico di Arcetri, Largo Enrico Fermi 5, I-50125 Firenze, Italy}

\author[0000-0002-2380-9801]{Anna de Graaff}
\affil{Leiden Observatory, Leiden University, P.O.Box 9513, NL-2300 AA Leiden, The Netherlands}

\author{Yasha Kaushal}
\affil{Department of Physics and Astronomy and PITT PACC, University of Pittsburgh, Pittsburgh, PA 15260, USA}

\author{Angelos Nersesian}
\affil{Sterrenkundig Observatorium, Universiteit Gent, Krijgslaan 281 S9, B-9000 Gent, Belgium}

\author{Jeffrey A.\ Newman}
\affil{Department of Physics and Astronomy and PITT PACC, University of Pittsburgh, Pittsburgh, PA 15260, USA}

\author[0000-0002-5027-0135]{Arjen van der Wel}
\affil{Max-Planck-Institut f\"ur Astronomie, K\"onigstuhl 17, D-69117, Heidelberg, Germany}
\affil{Sterrenkundig Observatorium, Universiteit Gent, Krijgslaan 281 S9, B-9000 Gent, Belgium}

\author[0000-0002-9665-0440]{Po-Feng Wu}
\affil{National Astronomical Observatory of Japan, Osawa 2-21-1, Mitaka, Tokyo 181-8588, Japan}

\begin{abstract}

The massive end of the gas-phase mass--metallicity relation (MZR) is a sensitive probe of active galactic nuclei (AGN) feedback that is a crucial but highly uncertain component of galaxy evolution models. In this paper, we extend the $z\sim0.7$ MZR by $\sim$0.5 dex up to log$(M_\star/\textrm{M}_\odot)\sim11.1$. We use extremely deep VLT VIMOS spectra from the Large Early Galaxy Astrophysics Census (LEGA-C) survey to measure metallicities for \numgal galaxies. The LEGA-C MZR matches the normalization of the $z\sim0.8$ DEEP2 MZR where they overlap, so we combine the two to create an MZR spanning from 9.3 to 11.1 log$(M_\star/\textrm{M}_\odot)$. The LEGA-C+DEEP2 MZR at $z\sim0.7$ is offset to slightly lower metallicities (0.05--0.13~dex) than the $z\sim0$ MZR, but it otherwise mirrors the established power law rise at low/intermediate stellar masses and asymptotic flattening at high stellar masses. We compare the LEGA-C+DEEP2 MZR to the MZR from two cosmological simulations (IllustrisTNG and SIMBA), which predict qualitatively different metallicity trends for high-mass galaxies. This comparison highlights that our extended MZR provides a crucial observational constraint for galaxy evolution models in a mass regime where the MZR is very sensitive to choices about the implementation of AGN feedback. 

\end{abstract}


\section{Introduction}
\label{sec:intro}

The mass--metallicity relation (MZR), the tight correlation between galaxy stellar mass and gas-phase oxygen abundance, is sensitive to the physical processes that govern the baryon cycle and shape galaxy evolution. Infalling gas fuels star formation and triggers the production of metals \citep{pagel1997}.  Feedback from star formation can then drive large scale outflows that eject gas and metals from a galaxy into the intergalactic or circumgalactic media (see \citealt{veilleux2005} and \citealt{tumlinson2017} and references therein), where it may eventually reaccrete onto the galaxy \citep{oppenheimer2010}.

At high masses, the steep fall off of the stellar mass function strongly implies a disruption of the baryon cycle and the quenching of star formation \citep{dimatteo2005}.  Many galaxy evolution models rely on some form of AGN feedback to shut off star formation \citep[e.g.,][]{springel2005, somerville2015}, but the dominant mechanism---such as thermal feedback that heats and ejects gas from the interstellar medium \citep{schaye2015} or radio-mode feedback that prevents cooling from the circumgalactic medium \citep{keres2005}---remains difficult to pin down with current observations.  However, differences in AGN feedback prescriptions in cosmological simulations result in significantly different shapes of the high mass end of the MZR at intermediate redshifts \citep{dave2017, torrey2019}, making it a key observational constraint for differentiating between AGN feedback mechanisms.

The MZR has been widely studied in the local universe \citep{tremonti2004, andrews2013, curti20}, at intermediate redshifts \citep[$z=0.5-1.5$; e.g.,][]{perezmontero2009, zahid2011, topping2021}, beyond cosmic noon with ground-based spectroscopic samples \citep[$z=1.5-3.5$, e.g.,][]{maiolino2008, shapley2017, sanders2018, sanders2020, topping2021, sanders2021}, and recently extending up to $z\sim8$ with the advent of JWST \citep{langeroodi2022, curti2023, nakajima2023}.  In the local universe, metallicity increases with stellar mass according to a power law as 12 + log(O/H)~$\propto M^{1/3}$ below a characteristic turnover mass ($\sim10^{10.25}\textrm{M}_\odot$), and then the MZR flattens as metallicities of the most massive galaxies asymptote to a constant value.  Generally, metallicity decreases at fixed stellar mass with increasing redshift.  The shape and normalization of the MZR evolve with redshift (e.g., \citealt{lilly2003}, \citealt{moustakas2011}, \citealt{sanders2020}, \citealt{sanders2021}), with the metallicities of lower mass galaxies evolving more dramatically across cosmic time \citep{zahid2013}.  However, above the turnover mass ($\log M_{\star}/M_{\odot} \gtrsim 10.25$), the MZR has only been reliably measured locally due to the difficulty of detecting the required emission lines in metal-rich galaxies at higher redshifts.

In this work, we take advantage of the ultra-deep spectra of a nearly mass-complete sample at $0.6<z<1.0$ obtained by the Large Early Galaxy Astrophysics Census \citep[LEGA-C][]{vanderwel2016, straatman2018} survey.  The primary science driver of the LEGA-C survey was the study of stellar populations and dynamics of massive galaxies at $z\sim0.8$, so the spectra have a much higher signal-to-noise ratio (S/N) per pixel than typical surveys of emission line galaxies.  LEGA-C's high S/N enables detection of even weak emission lines, significantly mitigating the observational bias towards objects with bright emission lines and systematically lower metallicities.  Despite the extremely long integration time required to achieve the high S/N per pixel, the LEGA-C survey is large enough to construct a statistically robust MZR up to $10^{11.1}\mathrm{M}_{\odot}$---high enough to constrain the effects of AGN feedback on the MZR.

The paper is structured as follows: \S 2 describes the LEGA-C data set, sample selection methods, and Bayesian metallicity estimation method; \S 3 presents the MZR at $z\sim0.7$, and \S 4 compares the LEGA-C MZR with cosmological hydrodynamic simulations. Throughout this paper we assume the following cosmology: H${_0}$ = 70 km s$^{-1}$ Mpc$^{-1}$, $\Omega_M$ = 0.3, and $\Omega_\Lambda$ = 0.7.

\section{Data and Analysis}
\label{sec:data_analysis}

\subsection{The LEGA-C Survey}

\begin{figure*}[htp]
    \centering
    \makebox[\textwidth][c]{\includegraphics[width= \textwidth]{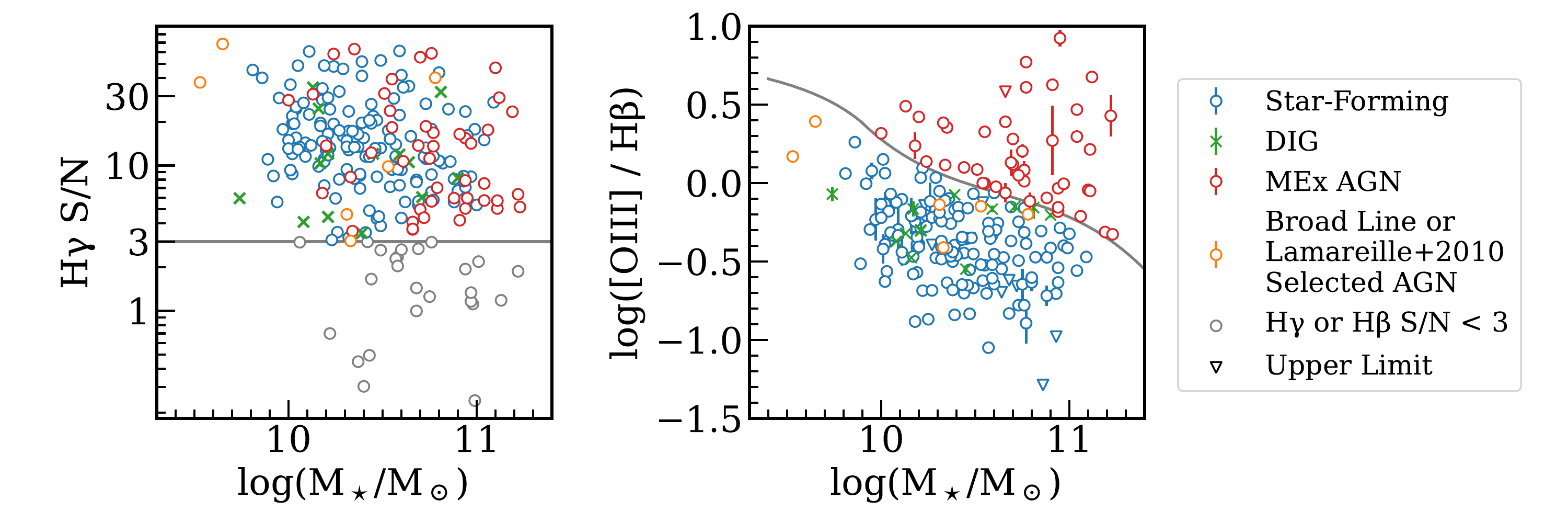}}
    \caption{Left panel: \hg\ S/N for non-quiescent LEGA-C objects as a function of stellar mass. Gray points are those with S/N $<$ 3, which are excluded from the sample. The remaining points follow the same color scheme as the right panel, described below. We do not find a trend of \hg\ S/N with stellar mass. Only a small fraction of these objects have \hg\ S/N $<$ 3, which is required for dereddening. Right panel: The same sample on the mass--excitation (MEx) plot of \citet{juneau2011}, which plots R3 against stellar mass and separates star-forming galaxies and AGN based on their line ratios. The red points show objects that are determined to be MEx AGN by the cut of \citet{juneau2011}. Orange points show objects determined to be other forms of AGN, such as broad-line, radio, IR, and through the ``blue diagram'' of \citet{lamareille2009}. Green crosses show objects with line fluxes dominated by DIG, and blue points show star-forming objects with line fluxes dominated by HII regions.}
    \label{fig:selection}
\end{figure*}

The LEGA-C survey used a \textit{K}-band selection to target a nearly mass-selected sample of 3,741 high-mass ($M_\star \gtrsim 10^{10}$ M$_\odot$) galaxies with existing Ultra-VISTA photometry at $z=0.6-1$. The spectra were obtained with 20 hours of integration time each on the Very Large Telescope (VLT) with the VIsible MultiObject Spectrograph \citep[VIMOS][]{lefevre2003} and resulted in spectra over the wavelength range 5800$-$9500\AA\ with an average S/N of 20 \AA$^{-1}$. Details of the survey design and sample selection can be found in \citet{vanderwel2016}, \citet{straatman2018}, and \citet{vanderwel2021}. LEGA-C's \textit{K}-band selection and high S/N is well-suited for measuring metallicity even for objects with weak emission lines and significant stellar Balmer absorption lines that would otherwise render the measurement of weak Balmer lines challenging. This opens up novel parameter space including the high mass, high metallicity, and low specific star formation rate (sSFR) regime.

We adopt the stellar masses and line flux measurements from the LEGA-C catalog (DR3; \citealt{vanderwel21}). Stellar masses were determined from UltraVISTA photometry (\citealt{muzzin2013}, \citealt{muzzin2013b}) using the FAST code \citep{kriek2009} and \citet{bruzual2003} stellar population libraries, assuming a \citet{chabrier2003} initial mass function and \citet{calzetti2000} dust extinction law, with an exponentially declining star formation rate (SFR). If we had chosen to adopt the MAGPHYS stellar masses (\citealt{dacunha2008}, as measured for LEGA-C galaxies by \citealt{degraaff2021apj}), this would systematically increase stellar masses by $\sim0.2$~dex.

In order to properly account for absorption underlying the Balmer emission lines used in this analysis, the stellar continuum must be modeled and subtracted. The stellar continua were fit using pPXF (\citealt{cappellari2004}, \citealt{cappellari2012}, \citealt{cappellari2017}) with MILES empirical stellar population templates \citep{vazdekis2010}. For further details see \citet{vanderwel2016} and \citet{bezanson2018b}. We use emission line fluxes from continuum-subtracted spectra (v3.11) fit by \citet{maseda2021}, which used Platefit \citep{tremonti2004, brinchmann2004, gunawardhana2020} to simultaneously fit emission lines as Gaussians. These Gaussians were fixed to the LEGA-C spectroscopic redshifts but allowed a $\pm$300 km $\mathrm{s}^{-1}$ velocity offset between the Balmer lines and forbidden lines such as \oii. Uncertainties on the line fluxes were determined by fitting to fluxes perturbed by wavelength-dependent noise vectors. The standard deviation of these fits was combined with the uncertainty in the overall flux calibration to give a line flux uncertainty. 

We use four emission lines (\oii, \hg, \hb, and \oiii) to identify AGN, determine nebular extinction, and calculate metallicity. This analysis, detailed in Sec.~\ref{subsec:diagnostics}, combines those lines to determine the  R2, R3, R23, and O32 line ratios, defined as:
\begin{subequations}
\begin{equation}
    \mathrm{R2} = \frac{[\mathrm{O} \textsc{ii}]\lambda3727}{\mathrm{H}\beta},
\end{equation}
\begin{equation}
    \mathrm{R3} = \frac{[\mathrm{O} \textsc{iii}]\lambda5007}{\mathrm{H}\beta},
\end{equation}
\begin{equation}
    \mathrm{R23} = \frac{[\mathrm{O} \textsc{ii}]\lambda3727 + [\mathrm{O} \textsc{iii}]\lambda4959,\lambda5007}{\mathrm{H}\beta}\textrm{, and}
\end{equation}    
\begin{equation}
    \mathrm{O32} = \frac{[\mathrm{O} \textsc{iii}]\lambda5007}{[\mathrm{O} \textsc{ii}]\lambda3727}, 
\end{equation}
\label{eqn:line_ratios}
\end{subequations}
where \oii\ refers to the [O \textsc{ii}] doublet. 

\subsection{Sample Selection}
\label{subsubsec:sample_selection}

We require \oii\ and \oiii\ for our analysis, which are only simultaneously accessible for a small subset of the spectroscopic sample. The high spectral resolution of LEGA-C results in rather limited wavelength coverage. Of the $\sim3500$ galaxies in the primary LEGA-C sample, only 425 galaxies include both spectral features. This selection limits the redshift range of the sample analyzed to $0.568 < z < 0.894$. We further limit our sample to star-forming galaxies using UVJ rest-frame colors, adopting the \citet{muzzin2013} color--color cuts designed for the Ultra-VISTA data set, leaving 245 objects. Finally, because we utilize \hb\ and \hg\ to deredden line fluxes, we require a S/N $3\sigma$ in each line flux. Since our Bayesian metallicity estimation method naturally handles upper limits on \oii\ and \oiii\ line fluxes, we do not apply a S/N cut on them to avoid biasing our sample.  However, we visually inspect the spectra that fall into this category (18 objects), rejecting half of them for incomplete line coverage. This leaves 211 galaxies.

The left panel of Fig.~\ref{fig:selection} shows the \hg\ S/N as a function of stellar mass, with a horizontal line at S/N~$=$~3 showing our selection. The gray points are those with \hg\ S/N~$<$~3; the remaining points follow the same coloring convention as the right panel of the figure, described below. After we select star-forming galaxies, we are limited by our ability to measure the Balmer lines, because we do not place an S/N cut on \oii\ or \oiii. We do not find a trend of \hg\ S/N with mass, and we therefore reach a high level of completeness. We measure each emission line, on average, 92\% of the time. Since the LEGA-C survey was not designed to only target galaxies with the brightest emission lines, its extreme depth enables analysis of a much more complete set of star-forming galaxies than previous surveys. The high \hg\ detection fraction allows us to use the Balmer decrement to deredden spectra, instead of needing to assume an A$_\textrm{V}$ (see \citealt{savaglio2005}). For sample selection purposes only, we dereddened the line fluxes using the \hg/\hb\ Balmer decrement (e.g., equation A14 from \citealt{momcheva2013}). However, when we measure our metallicity, we simultaneously solve for 12 + log(O/H), $A_V$, and dereddened H$\beta$ flux, in both cases adopting a \citet{cardelli1989} extinction law with $R_V = 3.1$.

Gas-phase metallicities cannot be derived from the line ratios of AGN as the emission lines are not thermally excited, so they must be excluded. We are unable to use the Baldwin--Phillips--Terlevich diagnostic diagram \citep[BPT;][]{baldwin1981} to remove AGN from our sample as we do not have coverage of \nii\ or \ha. Instead, we identify AGN using the mass--excitation diagram \citep{juneau2011}, shown in the right panel of Fig.~\ref{fig:selection}. There are four populations on this panel. The red points denote objects that are shown to be mass--excitation AGN by \citet{juneau2011}. The orange points denote objects that are determined to be AGN by other means (mid-IR diagnostics, X-ray fluxes, radio emission, and broad emission lines; S.~Vervalcke priv.~communication). We also used cuts from the R3 vs.\ R2 diagram (see Section \ref{subsec:diagnostics} of \citealt{lamareille2010}) that only requires lines at the bluer end of the optical range to remove additional likely AGN (the LINER, Seyfert 2, and star-forming/Seyfert 2 classes). Finally, we remove galaxies whose emission is dominated by diffuse ionized gas (DIG) via the $\Sigma_{\mathrm{H}\alpha}$ cut suggested by \citet{zhang2017}, and retain galaxies with log($\Sigma_{\mathrm{H}\alpha} [\mathrm{erg} \, \mathrm{s}^{-1} \, \mathrm{kpc}^{-2}]$)~$< 39$. These are the objects shown as green crosses in Fig.~\ref{fig:selection}. For the purposes here, we calculate \ha\ via the intrinsic ratio between dereddened \hb\ and \ha. The removal of these objects has a minimal effect on the best-fit MZR. The blue points show galaxies that are star-forming and make up the final sample used in the rest of this paper. 

Of the 211 galaxies with coverage of the emission lines of interest, 52 are identified as AGN (orange and red points in Fig.~\ref{fig:selection}), and 159 are identified as star-forming. Of these 159 star-forming galaxies, 145 are found to have line fluxes dominated by HII regions. This latter group constituting the sample used in the rest of the analysis. The median redshift of our final sample is $z=0.70$, with stellar masses from log($M_\star/\textrm{M}_\odot) = 9.81$ to $11.1$. The median log(SFR [$\textrm{M}_\odot \, \mathrm{yr}^{-1}$]) = $1.1$, ranging from $-0.14$ to $1.89$. The median log(sSFR [$\mathrm{yr}^{-1}$]) = $-9.3$, ranging from $-10.4$ to $-8.6$.

\subsection{Emission Line Ratios}

\begin{figure}[htp!]
    \centering
    \includegraphics[width=9cm]{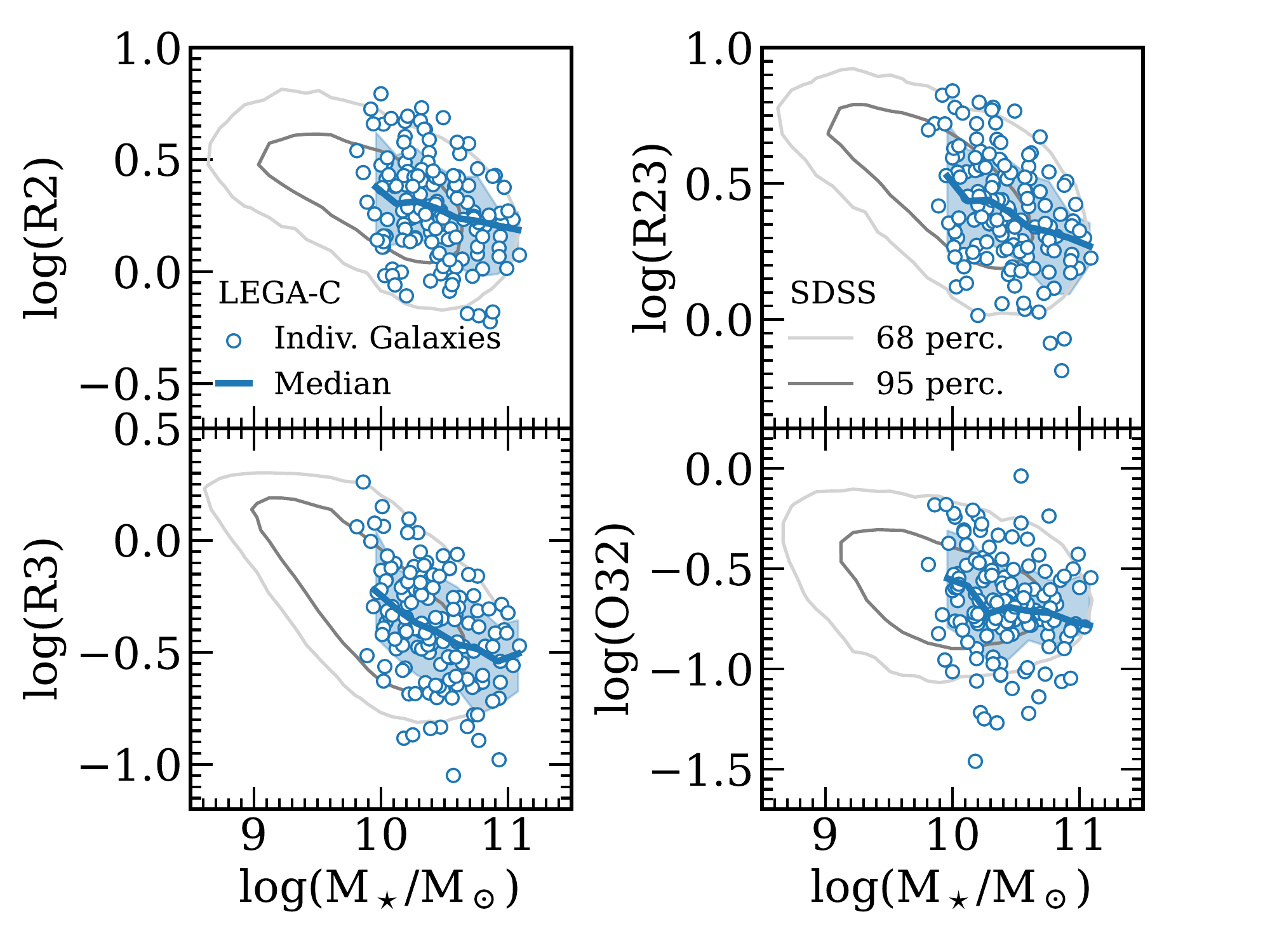}
    \caption{R2, R3, R23, and O32 versus stellar mass for the LEGA-C sample (blue circles with the running median and 16--84th percentiles as blue lines and bands, respectively) and SDSS (gray contours) samples. There is a slight offset in the LEGA-C data to higher line ratios for R2, R3, and R23, but not for O32.  We note that the median line ratio uncertainties are smaller than the symbol size if reddening uncertainties are not included. Note the smooth trend in line ratios and the lack of sharp features at log($M_\star/\textrm{M}_\odot) > 10.5$ that would be expected if AGN feedback dramatically alters the gas-phase metallicity at high mass.}
    \label{fig:line_ratios}
\end{figure}

\begin{table*}[htp]
\centering
\noindent\makebox[\linewidth]{}
\begin{tabular}{ccccccccc}
\toprule
ID & Mask & Redshift & log($\rm{M_\star / \textrm{M}_\odot}$) & $\frac{[O \textsc{ii}]\lambda3727}{H\beta}$ &  $\frac{[O \textsc{iii}]\lambda5007}{H\beta}$ & A$_V$ & 12+log(O/H) \\
\midrule
131198 & 1 & 0.726 & 10.94 & 1.47$\pm$0.07 & 0.23$\pm$0.02 & 1.91$_{-0.3}^{+0.29}$ & 8.78$_{-0.02}^{+0.02}$ \\
135418 & 1 & 0.74 & 11.04 & 1.8$\pm$0.07 & 0.27$\pm$0.03 & 1.6$_{-0.38}^{+0.39}$ & 8.76$_{-0.03}^{+0.03}$ \\
138799 & 1 & 0.718 & 10.04 & 2.98$\pm$0.13 & 0.67$\pm$0.03 & 1.46$_{-0.32}^{+0.31}$ & 8.65$_{-0.04}^{+0.03}$ \\
138923 & 1 & 0.757 & 10.19 & 2.78$\pm$0.1 & 0.39$\pm$0.03 & 1.37$_{-0.43}^{+0.42}$ & 8.71$_{-0.03}^{+0.03}$ \\
208640 & 1 & 0.702 & 10.73 & 1.94$\pm$0.09 & 0.56$\pm$0.03 & 0.43$_{-0.25}^{+0.3}$ & 8.68$_{-0.03}^{+0.03}$ \\
\bottomrule
\end{tabular}
\caption{A snippet of the results from our Bayesian metallicity estimation method. Columns, from left to right, are the galaxy's LEGA-C ID and slit mask, spectroscopic redshift, stellar mass, dereddened \oii\ and \oiii\ flux relative to \hb, nebular extinction A$_V$, and gas-phase metallicity, the latter four columns provide the inner 68 percentile highest probability density interval from the MCMC posterior distribution. The full table is available in machine-readable format.}
\label{table:results}
\end{table*}

It is useful to compare emission line ratios between studies as a first step, given that they are not affected by the choice of metallicity calibration. Fig.~\ref{fig:line_ratios} shows line ratios versus stellar mass for four common metallicity-sensitive diagnostics (R2, R3, R23, and O32) for LEGA-C galaxies whose line fluxes are dominated by HII regions. The LEGA-C line ratios are shown in blue circles with the blue line and shaded region indicating the running median and one-sigma scatter, respectively. Line flux ratios from the Sloan Digital Sky Survey (SDSS, \citealt{york2000}) MPA-JHU catalog (\citealt{kauffmann2003b}, \citealt{brinchmann2004}, \citealt{tremonti2004}) from the local universe are shown as gray contours. The LEGA-C trends generally agree with the SDSS data with a slight offset to higher R2, R3, and R23 in agreement with the more detailed analysis of LEGA-C emission line ratios of \citet{helton2022}. The similarity between the line ratios of LEGA-C and SDSS galaxies motivates our adoption of the \citet{curti2017} strong line metallicity calibration that is calibrated to direct method metallicities of stacked SDSS spectra.  Furthermore, smoothness of the trends and the absence of sharp features at log($\mathrm{M}_\star/\textrm{M}_\odot) > 10.5$ implies that the MZR should be smooth in this regime regardless of the metallicity calibration---in contrast to some simulation predictions with certain AGN feedback prescriptions (see Section \ref{sec:summary} for more discussion).

\subsection{Measuring Gas-Phase Metallicities}
\label{subsec:diagnostics}

We estimate metallicities using a Bayesian framework (see \citealt{wang2017}, \citealt{wang2019}) to account for covariance between the dust correction and metallicity.  We use the \texttt{emcee} package \citep{foreman-mackey2013} to run a Markov Chain Monte Carlo (MCMC) that samples combinations of metallicity, nebular dust extinction, and dereddened \hb\ flux ($f_\mathrm{H\beta}^{\mathrm{dered}}$), which can be used to generate \oii, \hg, \hb, and \oiii\ line fluxes given a metallicity calibration (\citet{curti2017} in our case) and assuming case B recombination for the Balmer line ratio.  We adopt a flat prior on metallicity (12~+~log(O/H) = [7.6, 9.3])\footnote{Formally, our metallicity range extends beyond the calibrated maximum metallicity of the \citet{curti2017} calibrations (12~+~log(O/H) = 8.85), so as to not impose an artificial ceiling on metallicity.  We treat objects with metallicities well-beyond this value with caution, which happens to be only a single object in our sample, whose inclusion does not affect the shape of the MZR or any of our main conclusions.} and on nebular extinction (A$_V$ = [0, 4]).  We use the Jeffreys' prior distribution for $f_\mathrm{H\beta}^{\mathrm{dered}}$ and limit its range to [0 to 10$^5$] in units of 10$^{-19}$ erg s$^{-1}$ cm$^{-2}$. The $\chi^2$ statistic for our MCMC is given by
\begin{equation}
    \chi^2 \propto \sum\limits_{i} \frac{(f_i^{\mathrm{dered}} - R_i \cdot f_{\mathrm{H\beta}}^{\mathrm{dered}})^2}{\sigma_{f_i^{\mathrm{dered}}}^2 + \sigma_{R_i}^2 (f_{\mathrm{H\beta}}^{\mathrm{dered}})^2},
\label{eqn:chisq}
\end{equation}
where $f^{\mathrm{dered}}_i$ is the dereddened line flux of emission line \textit{i}, $\sigma_{f_i^{\mathrm{dered}}}$ is its measured uncertainty, $R_i$ is the ratio of the dereddened flux of emission line $i$ to the dereddened H$\beta$ flux, and $\sigma_{R_i}$ is the intrinsic scatter in the $R_i$--metallicity relation from \citet{curti2017} (where $\sigma_{R_i}=0$ for the Balmer lines because we assume no uncertainty in the case B recombination \hb/\hg\ line ratio).

With our four emission lines (\oii, \hg, \hb, and \oiii), it is possible to use R2 and R3 as the line ratios of interest ($R_i$ in Equation \ref{eqn:chisq}) or R23 and O32.  We found similar results for both approaches, but the R2--R3 combination produced slightly smaller uncertainties, so we adopt those results as our fiducial measurements going forward.  The dereddened line fluxes, $\mathrm{A}_V$, and metallicities for individual galaxies are provided in Table~\ref{table:results}.

As a check on the Bayesian approach, we also calculated metallicities using the more common approach of first determining the reddening from the Balmer decrement and then estimating metallicity from the dereddened line flux ratios.  We found similar median estimated metallicities (average discrepancy of 0.03 dex), but underestimated uncertainties (average discrepancy of 0.05 dex) using the \citet{kk04} and \citet{curti2017} R23 calibrations (assuming the upper branch of R23 and after adjusting to the same calibration scale using the transformations from \citealt{teimoorinia2021}) compared to our Bayesian approach.

\section{Mass--Metallicity Relation at $\lowercase{z}\sim0.7$}
\label{sec:mzr}

\begin{figure*}[htp!]
    \centering
    \includegraphics[width=\textwidth]{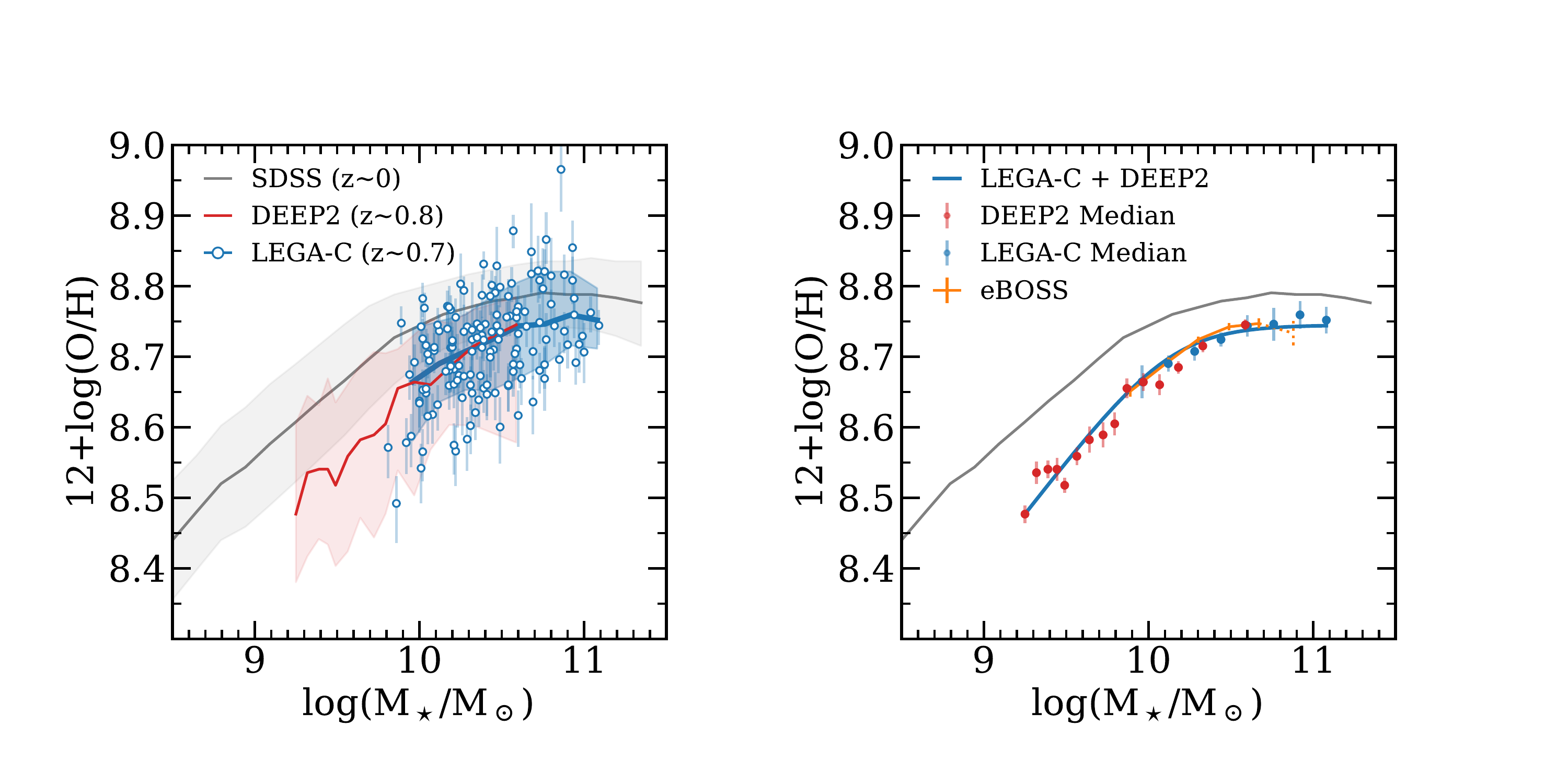}
    \caption{Left: The LEGA-C mass--metallicity relation with individual measurements shown as blue circles with error bars indicating the inner 68 percentile highest probability density interval; the median and the 16--84th percentile range are shown as a blue line and band, respectively. The $z\sim0$ SDSS MZR \citep{curti20} is shown in gray for comparison. The LEGA-C MZR shows a $-$0.05 dex offset from the SDSS MZR at the high mass end, and a $-$0.07 dex offset at the low mass end. The $z\sim0.8$ DEEP2 MZR \citep{zahid2011} (red line with red error band showing the 16--84th percentile region) agrees well with the LEGA-C MZR where they overlap in mass. Right: the LEGA-C and DEEP2 running median MZRs are shown as blue and red circles with error bars indicating the uncertainty on the median. The combined LEGA-C+DEEP2 MZR is shown as a blue line. The \citet{huang2019} MZR from stacks of eBOSS galaxies (orange) at $z\sim0.7$ is shown in orange with the most massive stellar mass bin indicated with a dotted line due to its low S/N (as noted by \citealt{huang2019}).  Overall, the eBOSS MZR agrees well with the LEGA-C+DEEP2 MZR where they overlap.}
    \label{fig:mzr}
\end{figure*}

Fig.~\ref{fig:mzr} shows the LEGA-C MZR in comparison to other $z \sim 0.7$ MZRs and the $z \sim 0$ SDSS MZR.  The left panel of Fig.~\ref{fig:mzr} shows individual LEGA-C galaxies as blue circles, the running median (with a window size of 0.5~dex) as a blue line, and the 16--84th percentile region as a blue band.  It also shows the \citet{zahid2011} MZR at $z\sim0.8$ (red) using data from DEEP2 (\citealt{davis2003}, \citealt{newman2013}) and the \citet{curti20} $z\sim0$ MZR for SDSS galaxies (gray).\footnote{For consistency, we decreased the MPA-JHU stellar masses \citep{kauffmann2003b, salim2007} adopted by \citet{curti20} by 0.2~dex to better match the normalization of the LEGA-C, DEEP2, and eBOSS stellar masses (see Fig.~2 of \citealt{zahid2011}), and we adjusted all metallicities to the \citet{curti2017} scale using the transformations from \citet{teimoorinia2021}.}  

The LEGA-C MZR matches the DEEP2 MZR almost perfectly in the mass range probed by both surveys (log$(M_\star/\textrm{M}_\odot) = 10.0$--$10.6$) and smoothly extends an additional 0.5~dex in stellar mass.  We find that the LEGA-C MZR has slightly smaller scatter (0.123 dex) than the DEEP2 MZR (0.225 dex).  The LEGA-C and DEEP2 objects show significant overlap in their sSFR distributions, suggesting that these two samples are probing similar populations of star-forming galaxies though at somewhat different mass ranges.  Given these similarities, we do a combined fit to  the LEGA-C and DEEP2 MZRs to form a single $z\sim0.7$ MZR that spans 1.7~dex in stellar mass (blue line in the right panel of Fig.~\ref{fig:mzr}).

The LEGA-C, DEEP2, and combined LEGA-C+DEEP2 MZRs agree well with the \citet{huang2019} $z\sim0.7$ MZR\footnote{Specifically, we show their $z=0.60$--$0.80$ MZR and not their MZR over their entire redshift range ($z=0.60$--$1.05$).} (orange line in the right panel of Fig.~\ref{fig:mzr}) that was constructed using stacked spectra from the eBOSS survey \citep{dawson2016}.  The slight turnover at the highest mass point could be due to AGN contamination in the stack due to the difficulty of identifying AGN in individual eBOSS spectra.  With the LEGA-C data, we found that including AGN would introduce a downturn in the MZR at the highest masses, hence the need for removing them.

To determine the best fit MZR across the full mass range by combining LEGA-C and DEEP2 MZR, we use the LMFIT package \citep{newville2014} and the functional form given in Equation 2 of \citet{curti20}:
\begin{equation}
    12 + \textrm{log(O/H)} = Z_0 -\frac{\gamma}{\beta} \textrm{log}(1 +(\frac{M}{M_0})^{-\beta}),
    \label{eqn:curti_fit}
\end{equation}
where $M_0$ is the characteristic turnover mass, $Z_0$ is the saturation metallicity that the MZR asymptotically approaches, $\gamma$ is the power law index of the MZR below $M_0$, and $\beta$ characterizes how quickly the power law begins to flatten.  We use a bin size of 0.16~dex to smooth over small-scale variations while retaining the large-scale trend.

The uncertainties on the combined fit were found by resampling the median metallicities according to their uncertainties and refitting, but the uncertainties on the fit are not shown for clarity because they are on the order of a few hundredths of a dex at all masses. The parameters of the best fit LEGA-C+DEEP2 $z\sim0.7$ MZR are $Z_0=8.74\pm0.01$, $\mathrm{log}(M_0 / \textrm{M}_\odot)=10.13\pm0.05$, $\gamma=0.30\pm0.02$, and $\beta=1.99\pm0.05$. The LEGA-C+DEEP2 MZR lies at an average $-$0.07 dex vertical offset from the SDSS MZR, with a $-$0.05~dex offset at the high mass end and a $-$0.13~dex offset at the low mass end. The values of $\mathrm{log}(M_0 / \textrm{M}_\odot)$ and $\gamma$ are within 1$\sigma$ of the SDSS MZR.

The offset of the LEGA-C+DEEP2 MZR from the $z\sim0$ MZR ($\sim -0.1$~dex below the turnover and $\sim -0.05$~dex above it) is significantly less than offsets found by previous studies at similar redshifts \citep[e.g.,][]{perezmontero2009,lamareille2009}. The offsets in these studies of $\sim-0.2$ dex could be due to a bias towards galaxies with stronger emission lines, their smaller sample sizes, or lower masses.  Both of the aforementioned studies used data from VVDS \citep{lefevre2005}, which also used the VIMOS spectrograph but before it was upgraded to have better red sensitivity.

Finally, we checked to see if accounting for the second-order dependence of metallicity on SFR (in addition to stellar mass) would further reduce the scatter in metallicity, but we did not find evidence for this so-called fundamental metallicity relation \citep[FMR;][]{mannucci2010, lara-lopez2010} in our LEGA-C sample. The FMR has been found to hold out to $z\sim3.3$ \citep{sanders2021}, but all MZRs at $z\sim0.7$ and beyond are at lower masses than LEGA-C.  The FMR is difficult to constrain at the high-mass end because there is little dynamic range in SFR, and the trend may reverse depending on the adopted metallicity calibration \citep{curti20}, so perhaps our finding is not too surprising.  Further complicating the issue is that LEGA-C does not cover \ha, which would provide the most robust short timescale SFR measurements. Given that the measured LEGA-C SFRs probe both longer timescales and different regions of the galaxy than metallicity indicators, we expect that this added scatter could erase additional correlations. We tested whether the scatter in the LEGA-C MZR correlates with other SFR indicators (e.g., UV+IR, spectral energy distribution fitting) but did not find evidence for a clear fundamental metallicity relation.

\section{Discussion and Conclusions}
\label{sec:summary}

\begin{figure}[htp!]
    \includegraphics[width=8cm]{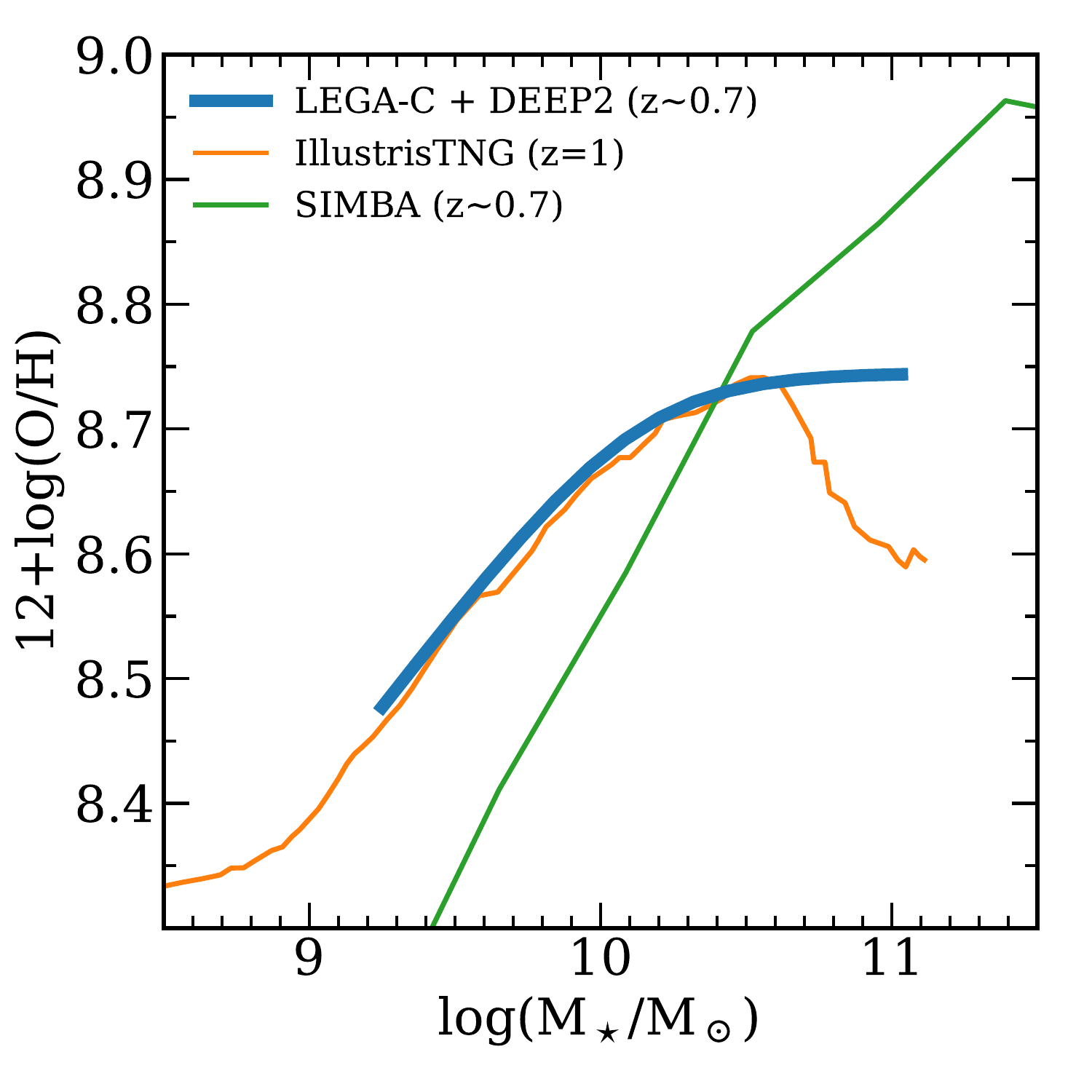}
    \caption{A comparison between our observed LEGA-C+DEEP2 MZR (blue) and MZRs from the IllustrisTNG simulation (orange; \citealt{torrey2019}) and the SIMBA simulation (green; \citealt{dave2019}) that shows the extreme sensitivity of the high mass end of the MZR to the implementation of AGN feedback in simulations. Given the large uncertainty in the overall normalization of the metallicity scale, we renormalized the simulated MZRs to match the LEGA-C+DEEP2 MZR at log($\mathrm{M}_\star/\textrm{M}_\odot)=10.4$. The IllustrisTNG MZR shows a significant decrease in metallicity above log($\mathrm{M}_\star/\textrm{M}_\odot)=10.6 $ in contrast to the LEGA-C MZR, likely due to AGN feedback ejecting too much metal-rich gas from central regions. The SIMBA MZR, on the other hand, continues to increase to log($\mathrm{M}_\star/\textrm{M}_\odot)=11$, which may be caused by insufficient metal-loading of winds, excessive recycling of winds, or radio-mode AGN feedback that results in closed-box-like metallicity evolution.}
    \label{fig:simulations}
\end{figure}

The LEGA-C MZR probes a very interesting stellar mass regime at $z\sim0.7$. The high-mass end of the MZR is important to constrain because of its sensitivity to AGN feedback in cosmological hydrodynamical simulations (\citealt{derossi2017}, \citealt{dave2019}, \citealt{torrey2019}), which are typically tuned to match the evolving stellar mass function (\citealt{weinberger2018}, \citealt{terrazas2020}, \citealt{zinger2020}). Because the exact implementation and strength of AGN feedback is hotly debated (\citealt{dave2017}, \citealt{torrey2019}), the observed MZR provides an important constraint and can be used to calibrate simulations.

In Fig.~\ref{fig:simulations} we compare the LEGA-C+DEEP2 MZR (blue) to the MZRs from the IllustrisTNG (orange; \citealt{torrey2019}) and SIMBA (green; \citealt{dave2019}) simulations at similar redshifts. We focus mainly on the difference in the shape of the MZRs due to uncertainties in the overall normalization of stellar masses, metallicities, and stellar yields, so we normalized the metallicities of the IllustrisTNG and SIMBA MZRs to match the LEGA-C+DEEP2 MZR at log($\mathrm{M}_\star/\textrm{M}_\odot)=10.4$. The IllustrisTNG MZR is shifted upwards 0.02 dex, the SIMBA MZR is shifted upwards 0.15 dex. The LEGA-C+DEEP2 MZR flattens at $\mathrm{log(O/H)}\sim8.75$ and qualitatively disagrees with both simulations at the high-mass end---the range of the MZR where LEGA-C becomes crucial. The SIMBA MZR uses an sSFR cut of $sSFR > -0.38$Gyr$^{-1}$ at $z=0.7$. The IllustrisTNG MZR uses mass-weighted metallicity as opposed to SFR-weighted metallicity to avoid highlighting exclusively star-forming gas \citep{torrey2019}. 

The IllustrisTNG simulation predicts a downturn in metallicity at log($\mathrm{M}_\star/\textrm{M}_\odot)=10.6$, in sharp contrast to the observed LEGA-C MZR, which flattens but does not decrease. This downturn, found at $z=1$ and $z=2$ but not at $z=0$, is driven by a transition from thermal to kinetic AGN feedback in the models \citep{weinberger2018} that removes large quantities of metal-rich gas from the centers of galaxies. Given the disagreement with the measured MZR, perhaps the ejective feedback within IllustrisTNG is removing too much metal-rich gas.

On the other hand, the SIMBA simulation predicts that the metallicity continues to increase up to log($\mathrm{M}_\star/\textrm{M}_\odot)=11$ before flattening. \citet{dave2017} speculate that the precursor of SIMBA simulation (i.e., MUFASA) may underestimate the metal-loading of winds or overestimate wind recycling in high-mass galaxies, and this problem may also be present in the SIMBA simulations. The median SIMBA MZR at $z\sim0.7$ (shown here) is robust to the SFR cut used to select star-forming galaxies.  The SIMBA MZR is in closer agreement with the LEGA-C+DEEP2 MZR at the high-mass end than the MUFASA $z\sim1$ MZR due to SIMBA's on-the-fly dust model, suggesting that the treatment of dust in simulations is important for reproducing the observed MZR \citep{dave2019}. It is also possible that SIMBA's implementation of quenching by heating halo gas to stop accretion (mimicking radio-mode AGN feedback as opposed to the ejective AGN feedback of IllustrisTNG) may result in effectively closed-box evolution to high metallicities.

Measuring the massive end of the MZR at $z\sim0.7$ is a significant step towards understanding how AGN feedback affects galaxies. Further observation of LEGA-C galaxies with PFS \citep{takada2014, greene2022} to measure \ha\ and \nii\ would allow AGN removal through the classic BPT diagram, as well as calculations of metallicity through other calibrations. Additionally, constraining the high-mass end of the MZR at $z\sim1-2$ with PFS and MOONRISE \citep{maiolino2020} is a natural extension of this work and will help to more concretely define the evolution of the MZR at high masses.  At even higher redshifts, JWST can probe the metallicities of the galaxies that will evolve into high mass objects at low redshift.  Finally, more detailed comparisons to cosmological simulations and theory will help improve our understanding of AGN feedback. 

\section*{Acknowledgements}

We thank Xin Wang, Jakob Helton, Allison Strom, Ryan Sanders, and Alice Shapley for their valuable input that improved this work. FDE acknowledges support by the Science and Technology Facilities Council (STFC), from the ERC Advanced Grant 695671 ``QUENCH''.

\software{
LMFIT \citep{newville2014}, 
Astropy \citep{astropy2013}, 
NumPy \citep{harris2020}, 
emcee \citep{foreman-mackey2013}, 
Pandas \citep{reback2020pandas}.
}

\bibliographystyle{aa}
\bibliography{lega-c_metallicities.bib}
\end{document}